\documentclass[10pt]{article}
\usepackage{amsmath}
\usepackage{amssymb}
\usepackage{graphicx}

\makeatletter
\@addtoreset{equation}{section}
\makeatother

\renewcommand\thefigure{\thesection.\@arabic\c@figure}
\renewcommand\thetable{\thesection.\@arabic\c@table}

\newcommand{\mc}[1]{{\mathcal #1}}

\newcommand{\bb}[1]{{\mathbb #1}}

\def\G{{\mc G}}
\def\Q{{\mc Q}}

\def\rhob{\bar\rho}

\def\yo{y_{\text{opt}}}

\newcommand{\KWASEP}{K_{\lambda}(\rho_a,\rho_b)}

\newcommand{\dd}{d}

\newcommand{\Dr}{\xi}
\newcommand{\Drst}{\xi^{st}}
\newcommand{\rhoh}{\rho^H}

\begin{document}
\title{Fluctuations in the weakly asymmetric exclusion
process with open boundary conditions}
\author{
B. Derrida\thanks{Laboratoire de Physique Statistique,
Ecole Normale Sup\'erieure, 24 rue Lhomond, 75005 Paris, France;
emails \texttt{derrida@lps.ens.fr , enaud@lps.ens.fr}} ,
C. Enaud${}^{*}$,
 C. Landim\thanks{IMPA, Estrada Dona Castorina 110,
CEP 22460 Rio de Janeiro, Brasil and CNRS UMR 6085,
Universit\'e de Rouen, 76128 Mont Saint Aignan, France;
e-mail \texttt{landim@impa.br}} , and
S. Olla\thanks{Ceremade, UMR CNRS 7534,
    Universit\'e de Paris Dauphine,
    Place du Mar\'echal De Lattre De Tassigny
    75775 Paris Cedex 16 - France;
e-mail \texttt{olla@ceremade.dauphine.fr}}
}
\date{\today}
\maketitle
%



\begin{abstract}
We investigate the fluctuations around the average
density profile in the weakly asymmetric exclusion process
  with open boundaries in the steady state. We show that these
  fluctuations are given, in the macroscopic limit, by a centered  Gaussian
  field and we compute explicitly its covariance function. We use two
  approaches. The first method is dynamical and based
  on fluctuations around the hydrodynamic limit. We prove that the density 
fluctuations evolve
  macroscopically according to an autonomous stochastic equation,
  and we search for the stationary distribution of this evolution.
  The second approach, which is
  based on a representation of the steady state as a sum over paths, allows one to write the density fluctuations in
  the steady state as a sum over two independent processes, one of
  which is the derivative of a Brownian motion, the other one being
  related to a random path in a potential.
\end{abstract}
\vskip10pt
\noindent
{\bf Key words:} Exclusion process, stationary non-equilibrium states, fluctuations
\vskip10pt
\noindent
\thanks{We thank J. L. Lebowitz for useful discussions.}

\section{Introduction}

Non equilibrium systems such as systems in contact with two
thermostats at unequal temperatures or with two reservoirs at unequal
densities are known to exhibit long range correlations in their steady
state \cite{Slsdip}. These long range correlations have been
calculated from the microscopic dynamics only in very few cases: mainly in the case of the symmetric
exclusion process \cite{Spohn83} and of the asymmetric exclusion
process \cite{DEL2004}.

In the present paper we focus on the weakly asymmetric exclusion process
(WASEP)
(for which the bias scales as the inverse of the system size). We show
that the fluctuations of density are Gaussian and that, as for the symmetric case, the direct calculation of the two point
function by fluctuating hydrodynamics agrees with the expression derived
by expanding the large deviation function around the average density
profile. We also show that, as for the asymmetric case \cite{DEL2004}, the correlation
functions can be expressed in terms of two independent random processes.
\bigskip

The asymmetric exclusion process describes the stochastic evolution of
a system of particles on a one dimensional lattice of $L$ sites.  Each
site of the lattice is occupied by at most one particle at a given
time. Thus a configuration of the system is specified by a series of
occupation numbers $\{\eta_i\}_{i=1,\dots, L}$, where $\eta_i=1$ if
site $i$ is occupied by a particle, $\eta_i=0$ if site $i$ is empty.

At any given time, each particle independently attempts to jump to one
of its neighboring sites on the lattice with rates which depend on the
direction of the jump. The jump succeeds only if the target site is
empty, otherwise the particle doesn't move. We choose to scale the
time such that the hopping rate to a right neighbor is $1$, and we
note by $q$ the hopping rate in the opposite direction. Physically,
the asymmetry between left and right hopping rates mimics the effect
of some external field acting on the particles.

The first and last sites of the lattice, respectively labeled $1$ and
$L$, are in contact with reservoirs of density respectively $\rho_a$
and $\rho_b$. This can be achieved by adding, to the left of site
$1$, a site $0$ whose probability of being occupied by a particle is
kept constant to $\rho_a$ independently of the rest of the system,
and, to the right of site $L$, a site $L+1$ whose probability of being
occupied by a particle is kept constant to $\rho_b$ independently of
the rest of the system. Another equivalent way of expressing the
action of the left reservoir is to say that particles are added to the
site 1 at rate $\alpha=\rho_a$ when the first site is empty, and when
it is occupied, the particle is removed from the system at $\gamma=q
(1-\rho_a)$ (i.e.  a particle at site 1 attempts to jump to the left reservoir
(site 0) with a rate $q$ and the jump succeeds with a rate
$1-\rho_a$). In the same way, if  site $L$ is occupied, the
particle is removed at rate $\beta=1-\rho_b$, and if it is empty, a
particle is added at  rate $\gamma=q \rho_b$.

The generator $\mathbb{L}$ of the dynamics, acting on a given function $g$ of the
configuration of the system $\eta=\{\eta_i\}_{i=1,\dots, L}$ is given
by
\begin{eqnarray*} 
\mathbb{L}g(\eta) &=& \sum_{i=1}^{L-1} \eta_i
(1-\eta_{i+1})\big[g(\eta^{i,i+1})-g(\eta)\big] \\
&+& \sum_{i=2}^{L} q \eta_i (1-\eta_{i-1})
\big[g(\eta^{i-1,i})-g(\eta)\big] \nonumber \\
&+& \big[\rho_a (1-\eta_1)+q (1-\rho_a) \eta_1\big]
\big[g(\eta^1)-g(\eta)\big] \nonumber \\
&+& \big[q \rho_b (1-\eta_L)+ (1-\rho_b) \eta_L\big]
\big[g(\eta^L)-g(\eta)\big]\;, \nonumber
\end{eqnarray*}
where $\eta^{i,i+1}$ is the configuration obtained from $\eta$ by
exchanging the occupation numbers of sites $i$ and $i+1$
and  $\eta^{i}$ is obtained from $\eta$ by changing the occupation number
of site $i$.
\bigskip

In the following, we consider the weakly asymmetric exclusion
process, where the asymmetry $q$ scales with  the  system size $L$ 
by
\begin{align*} 
q&=1-\frac{\lambda}{L} \ \ \cdot
\end{align*}

Of particular interest are the macroscopic properties of the system in
the large $L$ limit. We focus here on the macroscopic density profiles
$\{\rho(x)\}$, $0\leq x\leq 1$, obtained by rescaling by $L^{-1}$ 
and smoothening the microscopic density profiles $\{\eta_i\}_{i=1,L}$. To
do that, we may for example divide the system in mesoscopic boxes of size $L_k$, with
$1 \ll L_k \ll L$. The macroscopic density $\rho_L(\frac{i}{L})$ is then
simply the number of particles in the box containing the site $i$,
divided by the size of the box. A more mathematical approach consists in
defining, for each size $L$,  the following distribution acting on  smooth test function 
\begin{equation} \label{def_rho}
\rho_L(t,x) = \frac 1L \sum_{i=1}^L \delta(x - i/L) \eta_i(L^2 t)
\end{equation}
In spite of the stochastic evolution of the system at the microscopic
scale, it is known that, as
$L\to \infty$, the macroscopic profile $\rho_L$ converges almost surely to a deterministic evolution $\rhoh(t,x)$
given by the solution of the hydrodynamic equations of the process
\cite{KLslips, MIPPshbmps, Slsdip}.  In the case of the weakly
asymmetric exclusion process, it has been proved \cite{MPS1989, G1987}
that this is given by the viscous Burgers equation:
\begin{equation} 
\label{burger}
\left\{
  \begin{array}{l}
\partial_t \rhoh =\partial^2_x \rhoh-\lambda 
(1-2 \rhoh)\partial_x \rhoh \;, \\
\rhoh(t,0)=\rho_a ,\ \  \rhoh(t,1)=\rho_b\; , \\
\rhoh(0,x) =\rho_0(x)
  \end{array}
\right.
\end{equation}
where $\rho_0(x)$ is the macroscopic limit profile at time $t=0$.

At a finer scale, the density profile $\rho_L$ has random fluctuations
of order $L^{-1/2}$ around the hydrodynamic trajectory $\rhoh$.  One
defines the density fluctuation field by
\begin{equation}
\label{ftdf}
\Dr_L(t,x) =\sqrt{L} \big[\rho_L(t,x)-\rhoh(t,x)\big] \ \ .
\end{equation} 
It is known \cite{MPS1989, HV02} that these fields have a well defined
large $L$ limit $\Dr(x,t)$ which is a generalized Ornstein-Uhlenbeck
process in the case of WASEP on an infinite lattice (for such system,
the parameter $L$ is not any more related to the size of the lattice,
which is infinite, but only to the rescaling of time and space, as
well as to the asymmetry rate).

For a finite system of size $L$, after a long time, the system
eventually reaches a steady state where the properties of the system
become time-independent. Except for particular values of the reservoir
densities and of the asymmetry parameter $q$ for which detailed balance is
satisfied \cite{ED2004}, this stationary state is
a non-equilibrium steady state, which differs from an equilibrium
state by the presence of a non-zero, site-independent average current
$j$
\begin{equation} 
\label{jeta}
j=\langle \eta_i (1-\eta_{i+1})\rangle-q \langle \eta_{i+1}
(1-\eta_i)\rangle
\end{equation}
(where the brackets $\langle . \rangle$ stands for the average with
respect to the steady state probability denoted by $\mu^L$).

The system presents almost surely an average macroscopic profile
$\{\rhob(x)\}$ such that, for any site $i$, one has in the large $L$
limit:
\begin{equation*}
\langle \eta_i \rangle \simeq  \rhob(\frac{i}{L}) \ \ .
\end{equation*}
Putting $\partial_t \rho$ to 0 into (\ref{burger}) gives the following
equation for the steady state average profile $\{ \rhob(x)\}$:
\begin{equation}
\label{rhob}
\left\{
\begin{array}{l}
\rhob'(x)=\lambda \rhob(x) (1-\rhob(x)) -J\;, \\
\rhob(0)=\rho_a \;,\\
\rhob(1)=\rho_b\;,
\end{array}
\right.
\end{equation}
where $J$ is an integration constant solution of
\begin{align*} 
\int_{\rho_a}^{\rho_b} \frac{\dd\rho}{\lambda \rho (1-\rho)-J}&=1 \ \ .
\end{align*}
$J$ is related to the steady state average current
$j$ (\ref{jeta})
by
\begin{align*} 
J&=\lim_{L\rightarrow \infty} L j \ \ \ .
\end{align*}
The steady state fluctuation field will be noted $\Drst$:
\begin{equation*}
\Drst(x) =\lim_{L \to \infty}\sqrt{L} \big[\rho_L(x)-\rhob(x)\big] \ \ 
\end{equation*}
with $\rho_L(x)$ defined as in (\ref{def_rho}).
\bigskip

In the stationary state the macroscopic limit of the density profile
coincides with the deterministic solution of the stationary equation
\eqref{rhob} with the steady current $J$ determined by the asymmetry
parameter $\lambda$ and the boundary conditions $\rho_a,
\rho_b$. Actually, there exist \cite{ED2004} explicit expressions for
$\rhob$, but we will not use them in this paper. 
Without loss of generality we assume that 
$$
\rho_a \;<\; \rho_b
$$
so that we have always $\rhob'(x) \ge 0$. The current $J$ can be
positive or negative depending on $\lambda$.

Our main result is that the macroscopic fluctuation field $\Drst(x)$
is a centered Gaussian field on $[0,1]$ with covariance given by
\begin{equation}
\label{eq:3}
\left< \Drst(x)\Drst(y) \right> = \chi(\rhob(x)) \delta(x-y) + f(x,y)
\end{equation}
where 
\begin{equation}
\label{chi}
\chi(\rho) = \rho(1-\rho)
\end{equation}
 and $f(x,y)$ is 
 given by
\begin{equation}
  \label{eq:4}
  f(x,y)= \frac{J\rhob'(x)\rhob'(y) \int_0^{x\wedge y}
\frac{\dd u }{\rhob'(u)}\int_{x\vee y}^1\frac{\dd v}{\rhob'(v)}}{\int_0^1
\frac{\dd u}{\rhob'(u)}} \ \ ,
\end{equation}
where $x\wedge y = \min\{x,y\}$ and $ x\vee y = \max\{x,y\}$.

In terms of the two points correlation function, this result implies
that when $x\neq y$
\begin{equation}
  \label{eq:2}
L \left( \left< \eta_{[Lx]} \eta_{[Ly]}\right> -
\left< \eta_{[Lx]}\right> \left<\eta_{[Ly]}\right>\right)  \mathop{\longrightarrow}_{L\to\infty}\
f(x,y) 
\end{equation}

\medskip
Some comments are in order.
\begin{itemize}
\item In the symmetric case ($\lambda = 0$) we
have $\rhob'(x) = \rho_b - \rho_a$ and $J= - (\rho_b- \rho_a)$, so that
$$
f(x,y) = {-(\rho_b- \rho_a)^2} {(x\wedge y)} {[1- (x\vee y)]} = -(\rho_b-
\rho_a)^2 (-\Delta)^{-1}(x,y)\; ,
$$ 
in agreement with the result of Spohn
on the 2-point correlation function \cite{Spohn83}.

\item In the equilibrium case $J=0$, the stationary fluctuation field is
just white noise on $[0,1]$, corresponding to the fact that the
stationary state will be given by a product measure.
\item The sign of $f(x,y)$ depends on the sign of the current $J$.  If
$\lambda$ is positive and large enough, $J > 0$ (the weak shock
regime) and $f(x,y) \ge 0$. In the other cases $J$ and $f$ are
negative.
\end{itemize}

We derive the result (\ref{eq:3},\ref{eq:4}) by two different
methods. The first approach is 
dynamical: we search for the stationary solutions of the macroscopic
fluctuations process.
The second approach is static and based on a representation (valid only
when $J(\rho_a-\rho_b)>0$, see \cite{ED2004}) of the weights in the steady
state as sums over paths which was used in \cite{ED2004} to calculate
the large deviation function of the
stationary measure. A priori there is no reason that the
correlation 
functions are simply related to the large deviation functional of the
density profile: density fluctuations describe variations of density
of order $\frac{1}{\sqrt{L}}$ whereas the large deviation functional
describes variations of order 1.
For the symmetric exclusion process, it was however shown
\cite{DLS2002jsp} that the
expression of correlation functions obtained by a direct calculation
can be recovered by expanding the large deviation functional around
the average profile. On the other hand, for the asymmetric exclusion
process, the large deviation functional is non-analytic close to the
average profile and there is no simple connection between the large
deviations and the two point correlations \cite{DLS2002}. One outcome of
the present work is that the fluctuations of the density are still given,
for the WASEP, as the expansion of the large deviation function around
the average profile.

Both methods rely on distinct representations of the kernel $f$. 
In the dynamical approach we proceed as follows.
Let $\mathcal L$ be the differential operator $\Delta + \lambda (1-
2\rhob(x) ) \nabla$ with Dirichlet boundary conditions on $[0,1]$. Since 
$\bar\rho$ is the solution of \eqref{rhob},
it is easy to show that $\mathcal L$ can be extended to a negative
self-adjoint operator   on $L^2(\rhob'(x)dx)$. The kernel of the
inverse operator $(-\mathcal L)^{-1}(x,y)$ can be calculated
explicitly (cf. Section \ref{sec-dyn})  and shown to satisfy
\begin{equation}
  \label{eq:5}
  f(x,y) = J \rhob'(x) \rhob'(y) (-\mathcal L)^{-1}(x,y)\;.
\end{equation}

In the static approach, one shows that the Gaussian field
$\Drst(x)$ can be decomposed as a sum of two \textbf{independent} processes
\begin{equation}
\label{eq:8}
  \Drst(x) = \sqrt{\frac{\chi(\rhob(x))}2} B'(x) + \frac 12 Y'(x)\;,
\end{equation}
where $B(x)$ is a standard Brownian motion such that
\begin{align}
\left<[B(x)-B(x')]^2\right>=|x-x'|
\end{align}
(i.e. $B'(x)$ is a standard
$\delta$-correlated white noise), while  $Y(x)$ is a centered Gaussian
process whose distribution is formally given by
\begin{equation}
  \label{eq:9}
 \dd\mathbb{Q}(\{Y\}) \propto \exp\left\{-\int_0^1 \dd x \left(\frac{-J
\rhob'(x)
      Y(x)^2}{2 \chi(\rhob(x))^2}+\frac{{Y}'(x)^2}{4 \chi(\rhob(x)) 
}\right)\right\} \mc{D}\big[\{Y\}\big]\;
\end{equation}
where $\mc{D}\big[\{Y\}\big]$ is the standard Feynman measure.
Writing the distribution of the centered Gaussian process $Y'(x)$ as
\begin{equation}
  \label{eq:11}
   \dd \mathbb{Q}(\{Y'\}) \propto \exp\left\{-\frac 12 \int_0^1\!\int_0^1 Y'(u) T(u,v)
   Y'(v) \;du\; dv\right\} \mc{D}[\{Y'\}]\;,
\end{equation}
we show in section 4 that its covariance can be written as
\begin{equation}
  \label{eq:10}
  \left< Y'(x) Y'(y) \right> = T^{-1}(x,y) = 
2 \chi(\rhob(x)) \delta (x-y) + 4  f(x,y) 
\end{equation}
where $f(x,y)$ can be explicitly calculated and shown to be given by
(\ref{eq:4}). The covariance (\ref{eq:3}) of the fluctuation field
$\Drst(x)$ follows from (\ref{eq:8}) and (\ref{eq:10}).

\section{Dynamical approach}
\label{sec-dyn}

Consider the time dependent fluctuation field $\Dr_L(t,x)$ defined by
(\ref{ftdf}). 
Under proper assumptions on the initial distribution, it is proved in
\cite{lmo} that the law of $\Dr_L$ converges, as $L\to \infty$, to the
solution $\xi(t,x)$ of the linear stochastic partial differential equation:
\begin{equation}
\label{eq:6}
\partial_t \xi = \partial_x^2 \xi - \lambda\, \partial_x
\{ [1-2\rhoh(t,x)] \xi \} - \partial_x
\left(\sqrt{2\chi(\rhoh(t,x))}\; W(t,x) \right) \;, 
\end{equation}
where  $W(t,x)$ is the standard space
time white noise, i.e., the Gaussian process on $ \mathbb R_+\times
(0,1)$ with covariance
\begin{equation*}
\left< W(t,q) W(t', q') \right> = \delta (q-q') \delta (t-t')\;,
\end{equation*}
and $\chi$ is given by (\ref{chi}).\\
Since $\xi(t,x)$ and $W(t,x)$ are distributions-valued processes, equation
(\ref{eq:6}) should be interpreted in the weak form: for  
any smooth test function $G$ with compact support in $(0,1)$, denoting by 
$\xi_t(G) = \int_0^1 G(x) \xi(t,x) dx$,  
\begin{eqnarray}
\label{eq:7}
\xi_t(G) - \xi_0(G) &=& \int_0^t \xi_s(G'') \, ds \;+\; \lambda \int_0^t
\xi_s \big( [1-2\rhoh(s, \cdot)] G' \big) \,ds  \\
&+& \int_0^t  \int_0^1  G'(x) \sqrt{2 \chi(\rhoh(s,x))}\; W(s,x)\dd s \dd
x\;.
\nonumber
\end{eqnarray}

To investigate the asymptotic behavior of $\xi_t$,
consider the weakly asymmetric exclusion process starting from a local
Gibbs state associated to the steady state density profile $\bar\rho$.
In this case, the solution of the hydrodynamic equation (\ref{burger})
is constant in time and equal to $\bar\rho$. In particular, the
density fluctuation field $\Dr_L$ converges, as $L\to\infty$, to
the Ornstein-Uhlenbeck process (\ref{eq:7}) with $\bar \rho (\cdot)$
in place of $\rhoh(s, \cdot)$. 
Let $\mc L$ be the differential operator defined by
\begin{align} \label{Ldiffop}
\mc L &=\Delta + \lambda (1-2 \bar \rho(x)) \nabla
\end{align}
 with Dirichlet conditions
at the boundary. An elementary computation shows that for any smooth
function $G$ with compact support on $(0,1)$  
\begin{equation}
\label{eq:12}
\xi_t(G) = \xi_0 (e^{t \mc L}G) + \int_0^t \int_0^1 
  \left(\nabla e^{(t-s) \mc L}G\right) (x) \, 
  \sqrt{2\chi(\bar \rho(x))}\; W(s, x)\dd s \dd x\;.
\end{equation}
Since we imposed Dirichlet boundary conditions for $\mc L$, $e^{t \mc
  L}G$ converges to $0$ as $t\to \infty$. The second term of the right
hand side of \eqref{eq:12} is a Gaussian variable with zero mean and
variance given by
\begin{equation*}
2 \int_0^t ds \int_0^1 dx \,
\left[\nabla \left(e^{(t-s) \mc L}G\right)(x)\right]^2 
\chi(\bar \rho(x))\;.
\end{equation*}
Integrating by parts the previous expression with respect to $x$, recalling the
explicit formula (\ref{Ldiffop}) for the operator $\mc L$ and equation (\ref{rhob})
for the stationary density profile, we obtain that the previous
integral is equal to
\begin{eqnarray*}
\!\!\!\!\!\!\!\!\!\!\!\!\! &&
\int_0^1 G(x)^2 \chi(\rhob(x)) \; dx \; -\;  
\int_0^1 (e^{t \mc L}G)(x)^2 \chi(\bar \rho(x)) \; dx \\
\!\!\!\!\!\!\!\!\!\!\!\!\! && \quad
+ \; 2 J \int_0^t  ds \int_0^1 dx \;
\left(e^{(t-s) \mc L}G (x) \right)^2 \; \bar \rho'(x) \;.
\end{eqnarray*}
Since $e^{t\mc L}G \to 0$ as $t\to \infty$, the second term of the
previous equation vanishes as $t\to \infty$.  On the other hand, since
$\bar \rho$ is solution of the stationary equation \eqref{rhob}, the
operator $\mc L$ is symmetric in $ L^2(\bar \rho'(x) dx)$. In
particular, the last term of the previous equation can be rewritten as
\begin{equation*}
2 J \int_0^t  ds \int_0^1 dx \; G(x)\,
[e^{2(t-s)\mc L}G] (x) \, \bar \rho'(x)
\end{equation*}
which converges, as $t\to \infty$, to
\begin{equation*}
J \int_0^1 G(x) [(- \mc L)^{-1} G](x) \; \bar \rho'(x)\; dx\;.
\end{equation*}

In conclusion, the generalized Ornstein-Uhlenbeck process $\xi_t$
converges, as $t\to\infty$, to a Gaussian field $\Drst$ with
covariance given by
\begin{eqnarray}
\label{eq:18}
\!\!\!\!\!\!\!\!\!\!\! &&
\left< \Drst(G), \Drst(F) \right>  \\
\!\!\!\!\!\!\!\!\!\!\! && \qquad 
=\;  \int_0^1 G(x) F(x) \chi(\bar
\rho(x)) \; dx 
\;+\;  J \int_0^1 G(x) [(- \mc L)^{-1} F](x) \; \bar \rho'(x)\; dx\;.
\nonumber
\end{eqnarray}
This Gaussian field is also the stationary state for the
Ornstein-Uhlenbeck process (\ref{eq:7}) and the limit of the density
fluctuation field under the stationary measure.

An elementary 
computation permits to rewrite the covariance as follows
\begin{eqnarray}
\label{eq:1}
\!\!\!\!\!\!\!\!\!\!\!\! &&
\left< \Drst(G), \Drst (F)\right> \;=\; \\
\!\!\!\!\!\!\!\!\!\!\!\! && \qquad
\int_0^1 G(x) \, F(x) \, \chi(\bar \rho(x)) \; dx
\;+\;  \int_0^1 dx \int_0^1 dy \, F(x) \, f(x,y) \, G(y)\;,
\nonumber
\end{eqnarray}
where $f$ is given by (\ref{eq:4}). Indeed, for $F=G$, the second term
on the right hand side of (\ref{eq:18}) can be written as
$$
 J \sup_{H} \Big\{ 2 \int_0^1 dx\, G(x)\, H(x) \, \rhob' (x)  \;-\;
\int_0^1 dx\, H (x) \, (- \mc L H) (x) \, \rhob' (x) \Big\} \; ,
$$
where the supremum is carried over all smooth functions $H$ 
vanishing at the boundary. 
Since $\bar\rho$ is the solution of the  equation (\ref{rhob}),
an integration by parts gives that 
$$
\int_0^1 dx\, H (x) \, (- \mc L H) (x) \, \rhob' (x) = \int_0^1 dx\,
[H' (x)]^2  \rhob' (x)\;.  
$$
On the other hand, since $H$ vanishes at the boundary, an
integration by parts permits to rewrite the linear term as
$$
2 \int_0^1 dx\, G(x)\, H(x) \, \rhob' (x) = - 2 \int_0^1 dx\, \Big(
\int_0^x dy \, \rhob' (y) G(y) - A \Big) H' (x) 
$$
for any constant $A$. In view of the previous two expressions, it
is not difficult to show that the supremum in $H$ is equal to
\begin{equation}
  \label{eq:cal1}
  \int_0^1 dx\, \frac 1{\rhob' (x)} 
\Big( \int_0^x dy \, \rhob' (y) G(y) - A \Big)^2
\end{equation}
with $A$ given by
$$
A\;=\; \frac 
{\int_0^1 dx\, \frac 1{\rhob' (x)} 
\int_0^x dy \, \rhob' (y) G(y)}
{\int_0^1 dx\, \frac 1{\rhob' (x)}} \;\cdot
$$
It remains to develop the square and to recall the factor $J$ in
front of the variational formula to obtain the second term on the
right hand side of (\ref{eq:1}) with $f$ given by (\ref{eq:4}).
\medskip

Notice that in the symmetric case (when $q=1$, i.e. $\lambda=0$), $\mc L$ is the usual Laplacian
$\Delta$, $\bar \rho'(x) = \rho_b - \rho_a = -J$, and \eqref{eq:18}
becomes
\begin{equation*}
\left< \Drst(G), \Drst(F) \right> =   \int_0^1 G(x) F(x) \chi(\bar \rho(x)) \; dx
  -  (\rho_b - \rho_a)^2 \int_0^1 G(x) [(-\Delta)^{-1} F](x) \; dx
\end{equation*}
in accordance with the covariance formula obtained by Spohn in
\cite{Spohn83}.

\section{Static Approach}

The idea of the derivation is similar to the one in the totally asymmetric
case \cite{DEL2004}. In \cite{ED2004}, under the assumption that
\begin{equation*} 
J (\rho_a-\rho_b) \;\geq\; 0\;,
\end{equation*}
(which holds, for example, when $\rho_a<\rho_b$ and $q>1$), 
it was showed that the probability
$\dd \bb{P}\big(\{\rho(x)\}\big)$ of observing a given macroscopic profile
$\rho(x)$ in the steady state can be written  in the large $L$ limit as
\begin{align} 
\label{Prho}
\dd \bb{P}(\{\rho\})&\sim \mc{D}[\{\rho\} ]
\int_{\{y\}} e^{-L \G(\{\rho\}, \{y\})}
\mc{D}[\{y\} ]
\end{align}
where the sum is over all positive continuous functions $\{y(x), 0\leq x
\leq 1 \}$. We will call such a function $\{y(x)\}$ a path.
The expression (\ref{Prho}) is written there as a path integral, but it
is nothing more than the large $L$ limit of a sum over discrete paths
$y(i/L)=\frac{n_i}{L}$, and the measure (\ref{Prho}) can be simply thought
as the weight of these discrete paths.
%

The function $\G(\{\rho(x)\}, \{y(x)\})$ was given in \cite{ED2004},
equation (3.22):
\begin{multline*} 
\G(\{\rho(x)\};\{y(x)\})= -\KWASEP +
     \vphantom{-y(0)\log\frac{1-\rho_a}{\rho_a}-y(1) \log
        \frac{\rho_b}{1-\rho_b}+
        \int_{0}^{1}dx \left[-\log \frac{\left|1-e^{-\lambda
y}\right|}{\lambda}+\rho \log \rho
        +(1-\rho)\log{(1-\rho)}   +(1-\rho+y')
        \log{(1-\rho+y')}+(\rho-y')\log{(\rho-y')} \right]}
   y(0)\log\frac{\rho_a}{1-\rho_a}+y(1) \log \frac{1-\rho_b}{\rho_b}
 \\
  +\int_{0}^{1}dx \left[-\log \frac{1-e^{-\lambda y}}{\lambda}+\rho \log
\rho
  +(1-\rho)\log{(1-\rho)}
  \vphantom{-\log \frac{1-e^{-\lambda y}}{\lambda}+\rho \log \rho
      +(1-\rho)\log{(1-\rho)}+(1-\rho+y')
      \log{(1-\rho+y')}+(\rho-y')\log{(\rho-y')}}
  \right. \\
   \left.
   \vphantom{-\log \frac{1-e^{-\lambda y}}{\lambda}+\rho \log \rho
    +(1-\rho)\log{(1-\rho)}+(1-\rho+y')
    \log{(1-\rho+y')}+(\rho-y')\log{(\rho-y')}}+(1-\rho+y')
    \log{(1-\rho+y')}+(\rho-y')\log{(\rho-y')}
   \right]
   \vphantom{-y(0)\log\frac{1-\rho_a}{\rho_a}-y(1) \log
     \frac{\rho_b}{1-\rho_b}+
     \int_{0}^{1}dx \left[-\log \frac{1-e^{-\lambda y}}{\lambda}+\rho
\log \rho
     +(1-\rho)\log{(1-\rho)}   +(1-\rho+y')
     \log{(1-\rho+y')}+(\rho-y')\log{(\rho-y')} \right]}
\end{multline*}
where $\KWASEP$ is a normalization constant.
The quantity $e^{-L \G(\{\rho\}, \{y\})}\mc{D}[\{y\}]\mc{D}[\{\rho\}]$
can be thought as the joint probability of the profile $\{\rho\}$ and
the path $\{y\}$.\\
One can rewrite (\ref{Prho}) as 
\begin{align*}
\dd \bb{P}(\{\rho\})&=\mc{D}[\{\rho\}]
\int_{\{y\}} r(\{\rho\}|\{y\}) \dd \bb{Q}(\{y\})
\end{align*}
where $\dd \bb{Q}(\{y\})$ is the probability measure of the positive walk $\{y\}$
\begin{equation*}
\dd \bb{Q}(\{y\})=\mc{D}[\{y\}]\int_{\{\rho\}} e^{-L \G(\{\rho\}, \{y\})}
\mc{D}[\{\rho\}] \ \sim  \ \mc{D}[\{y\}]
e^{-L\Q(\{y\})} 
\end{equation*}
with $\Q(\{y\})=\inf \limits
_{\{\rho(x)\}}\G(\{\rho\}, \{y\})$, i.e.
\begin{multline} 
\label{Qy}
\Q(\{y\})=-\KWASEP -\log 4+
     \vphantom{-y(0)\log\frac{1-\rho_a}{\rho_a}-y(1) \log
        \frac{\rho_b}{1-\rho_b}+
        \int_{0}^{1}\dd x \left[-\log \frac{\left|1-e^{-\lambda
y}\right|}{\lambda}+(1+y')
        \log{(1+y')}+(1-y')\log{(1-y')} \right]}
   y(0)\log\frac{\rho_a}{1-\rho_a}+y(1) \log \frac{1-\rho_b}{\rho_b}
 \\
  +\int_{0}^{1}\dd x \left[-\log \frac{1-e^{-\lambda y}}{\lambda}
  \vphantom{-\log \frac{1-e^{-\lambda y}}{\lambda}+(1+y')
      \log{(1+y')}+(1-y')\log{(1-y')}}
  \right. \\
   \left.
   \vphantom{-\log \frac{1-e^{-\lambda y}}{\lambda}+(1+y')
    \log{(1+y')}+(1-y')\log{(1-y')}}+(1+y')
    \log{(1+y')}+(1-y')\log{(1-y')}
   \right]
   \vphantom{-y(0)\log\frac{1-\rho_a}{\rho_a}-y(1) \log
     \frac{\rho_b}{1-\rho_b}+
     \int_{0}^{1}\dd x \left[-\log \frac{1-e^{-\lambda y}}{\lambda} +(1+y')
     \log{(1+y')}+(1-y')\log{(1-y')} \right]}
\end{multline}
(as for a given $y$, the optimal profile is 
$\rho_y=\frac{1+y'}{2}$)
and $r(\{\rho\}|\{y\})$ is the conditional probability of the profile
${\rho}$ given the path $\{y\}$ given by the Radon-Nikodym derivative
\begin{align*}
r(\{\rho\}| \{y\})&=\frac{e^{-L \G(\{\rho\}, \{y\})}
\mc{D}[\{y\}]}{\dd \bb{Q}(\{y\})}
\end{align*}
so
\begin{multline*}
\frac{\log\big(r(\{\rho\}| \{y\})\big)}{L}=-\log 4+
        \int_{0}^{1}\dd x \left\{-\rho \log \rho
-(1-\rho)\log{(1-\rho)}\vphantom{(\rho-y')\log{(\rho-y')}} \right. \\
+ (1+y') \log{(1+y')}+(1-y')\log{(1-y')}\\
 \left.  - (1-\rho+y') \log{(1-\rho+y')}-(\rho-y')\log{(\rho-y')}\right\}\\
\end{multline*}

The fluctuations of $\rho(x)$ in (\ref{Prho}) have thus
two contributions: one coming from  the 
choice of the path $y(x)$, and the other one from the randomness
of the profile 
$\rho(x)$ once the path $y(x)$ is chosen. We shall see that for small
fluctuations, these two
contributions are uncorrelated, leading to (\ref{eq:8}).

The optimal path $\yo(x)$, which maximizes
the expression (\ref{Qy}) of $\dd \bb{Q}(\{y\})$, is  solution of
\begin{align}
\yo'(0)&=2 \rho_a-1 \nonumber \\
\yo'(1)&=2 \rho_b -1 \nonumber  \\
\frac{2 \yo''}{1-\yo^{\prime 2}}&=-\frac{\lambda e^{-\lambda \yo}}{1-e^{-\lambda
\yo} } \ \ . \label{yo}
\end{align}
By expanding (\ref{Qy}) up to the second order, we get for  a path $y$ close to $\yo$:
\begin{align*}
y(x)&=\yo(x)+\delta y(x)
\end{align*}
that its probability measure is given by
\begin{align*}
\dd \bb{Q}(y)&\sim \exp \left({-L \int_0^1 \dd x \left[\frac{\lambda^2}{2} 
\frac{e^{-\lambda\yo}(\delta y)^2 }{(1-e^{-\lambda \yo})^2}+
\frac{(\delta y')^2}{1-\yo^{\prime 2}}\right]}\right)\mc{D}[\{y\}]
\end{align*}
As $\yo (x)>0$ and $\delta y\sim \frac{1}{\sqrt{L}}$, the condition that
\begin{align*} 
y(x)=\yo(x)+\delta y(x)&>0
\end{align*} 
is automatically satisfied.
\bigskip

The optimal density profile $\rho_y(x)$ for a given  $\{y\}$
(i.e. the one which maximizes  $r(\{\rho\}| \{y\})$) is given by
\begin{align} \label{rhoy}
\rho_y&=\frac{1+y'}{2}.
\end{align}
Given the fluctuation $\delta y(x) =y(x)-\yo(x)$ of the walk $\{y\}$, the
probability of a small fluctuation of density  $\delta
b(x)=\rho(x)-\rho_y(x)$ around $\rho_y$ is obtained by expanding
$r(\{\rho\}| \{y\})$ up to the second order
in $\delta b$ and $\delta y$:
\begin{align}
r(\{\rho\}| \{y\}) &\sim \exp \left({-L \int_0^1 \dd x (\delta b )^2
\frac{4}{1-\yo^{\prime 2}}}\right)\ \ \ \ .
\end{align}
 As $r(\{\rho\}|\{y\})$ does not depend of $\delta y $ at order $(\delta
b)^2$, the choice of $\delta b=\rho-\rho_y$ is
independent of the choice of the fluctuation of the path  $\delta y$.\\
The total density fluctuation $\delta \rho$ is then given by
\begin{align} 
\delta \rho &= \delta b+\rho_y -\rhob \\
&=\delta b+\frac{\delta y'}{2}\label{deltarho_t}
\end{align}
where we used (\ref{rhoy}).
Thus, by rewriting in (\ref{deltarho_t}) $\delta y=\frac{Y}{\sqrt{L}}$ and $\delta b=\dot{B}\frac{\sqrt{\rhob (1-\rhob)}}{\sqrt{2 L}}$, and  using 
(\ref{yo}), (\ref{rhoy}), (\ref{rhobyo}), and the fact that  
\begin{align} \label{rhobyo}
\rhob&=\frac{1+\yo'}{2}\ \ ,
\end{align}
 one gets (\ref{eq:8}).

\bigskip

We now make the link between the  small fluctuations
and the large deviation functional of the density. From (\ref{deltarho_t}),
we see that the probability of a
fluctuation $\delta \rho$ of order $\frac{1}{\sqrt{L}}$ around the
optimal profile $\rhob$ is  given by the sum over all path
fluctuations $\delta y$ of the probability of having a path fluctuation
$\delta y$ and a density fluctuation $\delta b=\delta \rho-\frac{\delta
y'}{2}$ around
$\rho_y$, i.e.
\begin{align} \label{Pdrho_tot}
\dd \bb{P}(\{\delta \rho\})&\sim \mc{D}[\{\delta \rho\} ]
\int_{\{\delta y\}} \mc{D}[\{\delta y\} ]e^{-L \int_0^1 \dd x \left[\frac{\lambda^2}{2} 
\frac{e^{-\lambda\yo}(\delta y)^2 }{(1-e^{-\lambda \yo})^2}+
\frac{(\delta y')^2+4 \left( \delta \rho -\frac{\delta y'}{2}\right)^2}{1-\yo^{\prime 2}}\right] } \ \ .
\end{align}
In fact, this expression follows directly from (\ref{Prho}).
Integrating it over $\delta y$ leads to a gaussian form for the density fluctuations $\delta \rho$ of order $\frac{1}{\sqrt{L}}$ which is another way of writing (\ref{deltarho_t}). On the other hand, considering (\ref{Pdrho_tot}) for $\delta \rho$ small but of order 1 (i.e. of order $L^0$ in $L$), and performing a saddle point evaluation over $\delta y$ leads to leading order in $\delta \rho$ to a deviation functional quadratic in $\delta \rho$ which is identical to the gaussian. The fact that this quadratic form of the large deviation functional (for $\delta \rho$ small but of order 1 in $L$) is equivalent to the expression of the gaussian fluctuations (for $\delta \rho \sim \frac{1}{\sqrt{L}}$) shows that for the WASEP the fluctuations of order $\frac{1}{\sqrt{L}}$ can be calculated by expanding the large deviation functional to leading order around the most likely profile. Mathematically, this is simply due to the fact that the saddle point calculation is exact when one deals with gaussian variables (here the $\delta y$). 

\section{Derivation of \eqref{eq:10}}
\label{sec:der}

Let us define 
\begin{equation*}
  a(x)= \frac{-\rhob'(x) J}{2 \chi(\rhob(x))^2}\;\cdot
\end{equation*}
Writing $Y(x)=Y(0)+\int_0^x Y'(s) \dd s$ and performing the Gaussian
integral over $Y(0)$ in (\ref{eq:9}), one obtains the following
expression for $T(u,v)$ in (\ref{eq:11}): 
\begin{equation} \label{Tuv}
T(u,v) = \frac{\delta(u-v)}{2\chi(\rhob(x))} + 2 \frac{\int_{u\vee
  v}^1 a(z) dz \int_0^{u\wedge v} a(z') dz'}{\int_0^1 a(z) dz}
\end{equation}
%
We show now that when one writes $T^{-1}(x,y)$ as in (\ref{eq:10}), one
gets
expression (\ref{eq:4}) for $f(x,y)$.\\
Firstly, the symmetry of $T^{-1}(x,y)$ implies the symmetry of $f(x,y)$:
\begin{align*}
f(x,y)&=f(y,x)\; .
\end{align*}
Then, by definition of the inverse of an operator, we have  
\begin{align} \label{invdef}
\delta(u-v)&=\int_0^1 T(u,t) T^{-1}(t,v) \dd t \ \ .
\end{align}
Inserting (\ref{Tuv}) and (\ref{eq:10}) in (\ref{invdef}) gives the following integral equation for $f(x,y)$:
\begin{multline} \label{fint}
\frac{ f(x,y)}{4 \chi(\rhob(x))}+\frac{1}{\int_0^1 a(t) \dd t} \left[ \int_x^1 \dd t
\int_t^1 a(z) \dd z \int_0^x a(z') \dd z' f (t,y) \right.\\
\left. +\int_0^x\dd t
\int_0^t a(z) \dd z \int_x^1 a(z') \dd z'
f(t,y)+\frac{\Theta(y-x)\chi(\rhob(y))}{2}
\int_y^1 a(z)\dd z \int_0^x a(z') \dd z' 
\right.\\
\left.
+\frac{\Theta(x-y)\chi(\rhob(y))}{2}
\int_x^1 a(z)\dd z \int_0^y a(z') \dd z'  \right]=0
\end{multline}
(\ref{fint}) implies that for $0<y<1$, one has 
\begin{equation} \label{fbound}
f(0,y) = f(1,y) = 0\ \ .
\end{equation}
Furthermore, when one applies the operator $\partial_x
\{\frac{\partial_x(.)}{a(x)}\}$ to (\ref{fint}), one gets the following differential equation:
\begin{equation}\label{fdiff} 
\partial_x \frac{\partial_x \left\{ \frac{f(x,y)}{4 \chi(\rhob(x))}
\right\}}{a(x)}-f(x,y)-\delta(x-y)\frac{\chi(\rhob(x))}{2}=0
\end{equation}
This equation can be written as
\begin{align} \label{eqf}
\partial_x^2 f(x,y)+\lambda \partial_x \left\{ (2 \rhob(x)-1)
f(x,y)\right\}+\delta(x-y) J \rhob'(x)=0
\end{align}
The $\delta(x-y)$ function simply means that the $\partial_x f(x,y)$ is
discontinuous in $x=y$, i.e. 
\begin{align}
\partial_1 f(x^-,x)-\partial_1 f(x^+,x)&= J \rhob'(x)
\end{align}
This differential equation with boundary condition (\ref{fbound}) is
equivalent to equation (\ref{eq:5}) and its solution is given by (\ref{eq:4}) (using equation (\ref{rhob}) and the
symmetry of $f(x,y)$ ) . 

\section{Conclusion}

We have proved that the density fluctuations  in the stationary state
of the weakly asymmetric exclusion process with open boundaries are
distributed like a Gaussian field.   
We have obtained a simple expression (\ref{eq:3}), \eqref{eq:4} of 
the two point function
of the weakly asymmetric exclusion process which extends the result of
Spohn \cite{Spohn83}. This correlation function is long ranged and is a
signature of the non locality of the large deviation function. As for the
symmetric case, expanding the large deviation functional of the density
around the optimal profile leads to the right expression.
Our results have also some similarity with those of the  totally asymmetric case
(TASEP) \cite{DEL2004}. There
too, the fluctuations of density can be written as a sum of two terms.
However for the TASEP one of the two processes (the Brownian excursion) is non Gaussian.
It is to be noted that this Brownian excursion is not the large
$\lambda$ limit of the process $Y(x)$ (\ref{eq:9}), and, more generally, the
correlation function (\ref{eq:3}) does not converge in the large
$\lambda$ limit to the correlation function of the totally asymmetric
case, meaning that the large $L$ limit and the large $\lambda$ limit
can't be inverted.

\end{document}